\documentclass{svjour2}

\smartqed

\usepackage{graphicx}

\journalname{Science Foundations of Physics}

\newcommand{\square}{{\kern1pt\vbox{\hrule height 1.2pt\hbox{\vrule width 1.2pt\hskip 3pt
   \vbox{\vskip 6pt}\hskip 3pt\vrule width 0.6pt}\hrule height 0.6pt}\kern1pt}}

\begin{document}

\title{Nonlocal Models of Cosmic Acceleration\thanks{This work was partially
supported by NSF grant PHY-1205591.}}

\author{R. P. Woodard}

\institute{R. P. Woodard \at
              Department of Physics \\
              University of Florida \\
              Gainesville, FL 32611 \\
              UNITED STATES \\
              Tel.: +1-352-392-8744 \\
              Fax: +1-352-846-0295 \\
              \email{woodard@phys.ufl.edu}}

\date{Received: date / Accepted: date}

\maketitle

\begin{abstract}
I review a class of nonlocally modified gravity models which were proposed to 
explain the current phase of cosmic acceleration without dark energy. Among the topics
considered are deriving causal and conserved field equations, adjusting the model to
make it support a given expansion history, why these models do not require an elaborate
screening mechanism to evade solar system tests, degrees of freedom and kinetic stability, 
and the negative verdict of structure formation. Although these simple models are not
consistent with data on the growth of cosmic structures many of their features are 
likely to carry over to more complicated models which are in better agreement with 
the data.
\end{abstract}

\section{Introduction}\label{intro}

In 1998 Hubble plots of distant Type Ia supernovae \cite{SNIA}
revealed that the universe began accelerating somewhat over six
billion years ago. Supernova data since then have continued to
support this conclusion \cite{newSN} and it has been confirmed by
other data sets \cite{data}. This triumph of observational cosmology
continues to challenge fundamental theorists \cite{reviews}.

The data are consistent with general relativity operating on a
critical energy density whose current composition is about 70\%
cosmological constant, with about 30\% nonrelativistic and small
amounts of relativistic matter \cite{Yun,Yunbook}. However,
fundamental theorists have a hard time understanding why the
cosmological constant should be much smaller than any other scale in
physics and why it should have the right magnitude to so recently
come into dominance \cite{Lambdarevs}. Scalar potential models
\cite{earlyquint,Paul} can be devised to reproduce the observed
expansion history \cite{TW1,scalrecon} but they are difficult to motivate. 
Quantum effects from a very light scalar have also been suggested 
\cite{Parker}.

Modifications of gravity have been considered \cite{Mark}. However,
generalizing the Hilbert Lagrangian from $R$ to $f(R)$ \cite{NO}
represents the only local, metric-based, generally coordinate
invariant and stable modification of gravity \cite{RPW}. Because the
only model within this class which exactly reproduces the
$\Lambda$CDM expansion history is general relativity with $f(R) = R
- 2 \Lambda$ \cite{nof(R)}, more general $f(R)$ models can deviate 
from observation even without considering perturbations. 

There is greater freedom to modify gravity if locality is abandoned
\cite{nonloc,SW,DEW}, but one should keep in mind Newton's famous dictum of
1692/3 \cite{IN}:
\begin{quote}
{\it that one body may act upon another at a distance thro' a
Vacuum, without the Mediation of any thing else, by and through
which their Action and Force may be conveyed from one to another, is
to me so great an Absurdity that I believe no Man who has in
philosophical Matters a competent Faculty of thinking can ever fall
into it.}
\end{quote}
The great man's misgivings have certainly stood the test of time:
more than three centuries of observation and experimentation
have failed to reveal a single exception to the paradigm of local,
second order field equations. Proposed theoretical exceptions, such as 
string field theory, tend to have extra degrees of freedom which carry 
negative energy \cite{EW}. 

My view is that nonlocal modifications of gravity cannot be fundamental.
If these models describe nature, I believe their nonlocal structures
must derive from the gravitational vacuum polarization of infrared gravitons
which were produced during the epoch of primordial inflation \cite{TWa}. 
This is a natural expectation because accelerated expansion creates a vast
ensemble of infrared gravitons --- it causes the tensor power spectrum 
\cite{Alexei} --- and because the large vacuum energy of primordial 
inflation provides a dimension three self-interaction \cite{Sasha,TWb}. 
Of course loops of massless particles can produce long range effects, as
witness confinement in quantum chromodynamics (QCD). Gravitons with zero 
vacuum energy have no impact on cosmology merely because the self-interaction 
has dimension five \cite{Steve}. However, a dimension three coupling between 
massless particles ought to engender stronger infrared effects than the 
dimension four couplings of CQD \cite{Gabriele}.

Quantum gravitational self-interactions during inflation are suppressed by 
the small ($\sim 10^{-10}$) loop-counting parameter of quantum gravity, 
but they also grow with the number of inflationary e-foldings \cite{TWc}.
The growth occurs because more and more infrared gravitons come into causal 
contact the longer inflation persists. One consequence of it is that the 
interactions become nonperturbatively strong during a sufficiently prolonged 
period of inflation. That is frustrating because it means some sort of 
resummation technique must be devised in order to gain quantitative control 
over late time cosmology. This task is not hopeless --- it can be done for
scalar models \cite{SY,TW2} --- but it has not been accomplished yet for
quantum gravity \cite{MW}. In the absence of such a nonperturbative 
resummation technique the approach I shall describe here is the purely 
phenomenological one of proposing a plausible set of nonlocal field 
equations and examining their properties. However, it is well to note
that models which might eventually be derived from inflationary vacuum
polarization must possess two features which would be forbidden in 
fundamental theory:
\begin{itemize}
\item{There is an initial value surface corresponding to inflation's end; and}
\item{Because the putative effect derives from cosmological-scale gravitons 
it is completely natural that there should no change in gravity on small scales.}
\end{itemize}

The model I will describe in this article was developed with Stanley Deser. 
We change the gravitational Lagrangian from that of general relativity 
$\mathcal{L} = R\sqrt{-g}/16\pi G$ by the addition of a nonlocal term of the 
form \cite{ourmodel},
\begin{equation}
\Delta \mathcal{L} \equiv \frac1{16 \pi G} \, R \sqrt{-g} \times
f\Bigl(\frac1{\square} R\Bigr) \; . \label{DL2}
\end{equation}
Here $\square \equiv (-g)^{-\frac12} \partial_{\mu} [\sqrt{-g} \,
g^{\mu\nu} \partial_{\nu}]$ is the scalar d`Alembertian and we
define its inverse with retarded boundary conditions, which make
$\frac1{\square} R$ and its first time derivative vanish at the initial
time \cite{ourmodel}. (Note the appearance of an initial time!) In 
addition to simplicity, the great advantage of this class of 
models is to provide a natural delay for the onset of cosmic acceleration: 
because the Ricci scalar $R$ vanishes during radiation domination, 
$\frac1{\square} R$ cannot begin to grow until after the onset of matter 
domination, and then its growth is only logarithmic because of the 
inverse differential operator.

The model is defined by its field equations which take the form,
$G_{\mu\nu} + \Delta G_{\mu\nu} = 8 \pi G T_{\mu\nu}$. Section 
\ref{causcon} explains how to derive conserved field equations that 
are nevertheless causal. Section \ref{recon} describes how to
choose the function $f(X)$ in expression (\ref{DL2}) to make the model
reproduce the $\Lambda$CDM expansion history with zero cosmological
constant. This construction only determines the shape of $f(X)$ for $X 
< 0$, leaving its behavior for $X > 0$ completely free. Section \ref{screen} 
exploits this freedom to avoid any deviation from general relativity 
inside gravitationally bound systems. (Note the absence of small scale 
modifications, as foretold above.) Section \ref{DoF} demonstrates that 
the model possesses the same gravitational degrees of freedom and initial 
value constraints as general relativity, and that subsequent evolution never 
changes kinetic energies from positive to negative. Like all modified 
gravity theories, nonlocal cosmology can be differentiated from general 
relativity with dark energy by how it affects structure formation \cite{Yun2}. 
Although initial studies revealed no disagreement with the data \cite{Tomi,Park} 
a more recent analysis by Dodelson and Park \cite{Scott} has shown that the 
best current data heavily favors general relativity over the model (\ref{DL2}). 
That crucially important work is reviewed in section \ref{struc}. In section 
\ref{outlook} I argue that models involving a nonlocal invariant other than 
$\frac1{\square} R$ can likely be devised which meet the requirements 
of structure formation better than general relativity.

\section{Causality and Conservation from Nonlocal Actions}\label{causcon}

In this section we work out how our nonlocal modifications (\ref{DL2}) 
change the Einstein equations. These changes $\Delta G_{\mu\nu}[g](x)$ 
must be {\it conserved}, that is their covariant divergence must vanish,
$D^{\nu} \Delta G_{\mu\nu} = 0$. That follows automatically from varying 
a diffeomorphism invariant action, whether the Lagrangian is local or nonlocal. 

The nonlocal corrections we seek must also be {\it causal}. That is, the 
variation of $\Delta G_{\mu\nu}[g](x)$ with respect to fields at some 
point $y^{\mu}$, 
\begin{equation}
\frac{\delta {\Delta G}_{\mu\nu}[g](x)}{\delta g_{\rho\sigma}(y)} \; ,
\end{equation}
must vanish unless $y^{\mu}$ is on or within the past light-cone of
$x^{\mu}$. That cannot follow from the usual effective action by a simple 
symmetry argument. Consider the self-mass-squared correction to a scalar 
field $\varphi(x)$ in flat space,
\begin{equation}
\Delta \Gamma[\varphi] = -\frac12 \int d^4x \int d^4x' \varphi(x)
M^2(x;x') \varphi(x') \; .
\end{equation}
The scalar self-mass-squared is symmetric under interchange,
$M^2(x;x') = M^2(x';x)$, but even if it vanished for $x^{\prime\mu}$
outside the past light-cone of $x^{\mu}$, the effective field equations
would still be acausal because the variation affects both the field 
$\varphi(x)$ in the future and the field $\varphi(x')$ in the past,
\begin{equation}
\frac{\delta {\Delta \Gamma}[\varphi]}{\delta \varphi(y)} = -\frac12 \int 
d^4x \Bigl\{ M^2(y;x) + M^2(x;y) \Bigr\} \varphi(x) \; .
\end{equation}

One gets causal effective field equations from quantum field theory by
employing the Schwinger-Keldysh formalism \cite{SK}. Because we are 
just treating nonlocal models phenomenologically we shall instead enforce 
causality by resorting to two partial integration tricks. First, consider 
some functional of the metric times the variation of the nonlocal factor
$\frac1{\square} R$ \cite{SW},
\begin{eqnarray}
\lefteqn{\sqrt{-g} \, F[g] \frac{\delta}{\delta g^{\mu\nu}} \Bigl(
\frac1{\square} R\Bigr) = \sqrt{-g} \, F[g] \Biggl\{-\frac1{\square} 
\frac{\delta \square}{\delta g^{\mu\nu}} \frac1{\square} R + \frac1{\square} 
\frac{\delta R}{\delta g^{\mu\nu}} \Biggr\} \; ,} \label{init2} \\
& & \hspace{2.5cm} \longrightarrow -\sqrt{-g} \, \Bigl(\frac1{\square} F[g]\Bigr) 
\frac{\delta \square}{\delta g^{\mu\nu}} \frac1{\square} R + 
\sqrt{-g} \, \frac{\delta R}{\delta g^{\mu\nu}}
\frac1{\square} F[g] \; . \qquad \label{trick2}
\end{eqnarray}
A similar (illegal) partial integration is needed to vary $\square$ \cite{SW},
\begin{eqnarray}
\lefteqn{-\sqrt{-g} \Bigl(\frac1{\square} F[g]\Bigr) \frac{\delta \square}{
\delta g^{\mu\nu}} \Bigl(\frac1{\square} R\Bigr) \longrightarrow
\sqrt{-g} \, \Biggl\{ -\frac12 g_{\mu\nu} R \Bigl(\frac1{\square} F[g]\Bigr) } 
\nonumber \\
& & \hspace{2cm} -\frac12 g_{\mu\nu} \Bigl(\frac1{\square} R
\Bigr)^{,\rho} \Bigl(\frac1{\square} F[g] \Bigr)_{,\rho} +
\Bigl(\frac1{\square} R \Bigr)_{(,\mu} \Bigl(\frac1{\square} F[g]
\Bigr)_{,\nu)} \Biggr\} \; . \qquad \label{trick3}
\end{eqnarray}
Of course it is not correct to go from (\ref{init2}) to (\ref{trick2}), nor
from left to right in (\ref{trick3}), but one does get causal field equations 
this way provided $\frac1{\square}$ is always defined with retarded boundary 
conditions. Conservation is preserved because that depends only upon the 
relation $\square \times \frac1{\square} = 1$, which is valid whether 
$\frac1{\square}$ is retarded or advanced.

We should recall as well the familiar rule for varying the Ricci scalar,
\begin{equation}
\frac{\delta R(y)}{\delta g^{\mu\nu}(x)} = \Bigl[ R_{\mu\nu} - D_{\mu}
D_{\nu} + g_{\mu\nu} \square\Bigr] \delta^4(y-x) \; . \label{trick1}
\end{equation}
With this and our two partial integration tricks (\ref{trick2}-\ref{trick3})
it is straightforward to work out how $\Delta \mathcal{L}$ changes the
field equations. There are three kinds of terms. The first derive from varying 
the local factors of $R\sqrt{-g}$,
\begin{equation}
{\Delta G}^{a}_{\mu\nu} = \Bigl[R_{\mu\nu} - \frac12 g_{\mu\nu} R - D_{\mu} 
D_{\nu} + g_{\mu\nu} \square\Bigr] f\Bigl(\frac1{\square} R\Bigr) \; . 
\label{var2a}
\end{equation}
The second kind come from varying the $R$ inside the function
$f(\frac1{\square} R)$,
\begin{equation}
{\Delta G}^{b}_{\mu\nu} = \Bigl[R_{\mu\nu} - D_{\mu} D_{\nu} + g_{\mu\nu} 
\square\Bigr] \frac1{\square} \Biggl[ R f'\Bigl(\frac1{\square} R\Bigr) 
\Biggr] \; . \label{var2b}
\end{equation}
And the final kind comes from varying the $\frac1{\square}$,
\begin{eqnarray}
\lefteqn{{\Delta G}^{c}_{\mu\nu} =-\frac12 g_{\mu\nu} R \frac1{\square}\Biggl[ 
R f'\Bigl( \frac1{\square} R\Bigr) \Biggr] -\frac12 g_{\mu\nu} \Bigl(\frac1{
\square} R \Bigr)^{,\rho} \Biggl( \frac1{\square} \Biggl[ R f'\Bigl(\frac1{
\square} R \Bigr) \Biggr] \Biggr)_{,\rho} } \nonumber\\
& & \hspace{5cm} + \Bigl(\frac1{\square} R\Bigr)_{(,\mu} \Biggl( 
\frac1{\square} \Biggl[ R f'\Bigl( \frac1{\square} R \Bigr) \Biggr] 
\Biggr)_{,\nu)} \; . \qquad \label{var3}
\end{eqnarray}

Each of the three terms (\ref{var2a}), (\ref{var2b}) and (\ref{var3}) is manifestly
causal if we interpret the factors of $\frac1{\square}$ as retarded Green's functions.
To see that they are also conserved, note that acting $D^{\nu}$ on each term gives,
\begin{eqnarray}
D^{\nu} {\Delta G}^{a}_{\mu\nu} & = & -\frac12 R \partial_{\mu}
f\Bigl(\frac1{\square} R\Bigr) \; , \\
D^{\nu} {\Delta G}^{b}_{\mu\nu} & = & \frac12 R_{,\mu} \frac1{\square} 
\Biggl[ R f'\Bigl( \frac1{\square} R\Bigr)\Biggr] \; , \\
D^{\nu} {\Delta G}^{c}_{\mu\nu} & = & -\frac12 R_{,\mu} \frac1{\square} 
\Biggl[R f'\Bigl(\frac1{\square} R\Bigr)\Biggr] + \frac12 \Bigl(\frac1{
\square} R \Bigr)_{,\mu} R f'\Bigl(\frac1{\square} R\Bigr) \; . 
\end{eqnarray}
The nonlocal addition to the Einstein tensor is,
\begin{eqnarray}
\lefteqn{\Delta G_{\mu\nu} = \Bigl[ R_{\mu\nu} - \frac12 g_{\mu\nu} R + 
g_{\mu\nu} \square - D_{\mu} D_{\nu} \Bigr] \Biggl\{ f\Bigl(\frac1{\square} 
R\Bigr) + \frac1{\square} \Biggl[ R f'\Bigl(\frac1{\square} R\Bigr)\Biggr] 
\Biggl\} } \nonumber \\
& & \hspace{2cm} + \Bigl[\delta_{\mu}^{~(\rho} \delta_{\nu}^{~\sigma)} -
\frac12 g_{\mu\nu} g^{\rho\sigma}\Bigr] \Bigl(\frac1{\square} R
\Bigr)_{,\rho} \Biggl(\frac1{\square} \Biggl[ R f'\Bigl(\frac1{\square} R
\Bigr) \Biggr] \Biggr)_{,\sigma} \; . \qquad \label{mod2}
\end{eqnarray}
The term $G_{\mu\nu} \{f(\frac1{\square} R) + \frac1{\square} [R f'(\frac1{\square} 
R)] \}$ could be viewed as a sort of time-varying Newton constant, with the remaining 
terms there to enforce conservation.

\section{Reconstruction: $\Lambda$CDM without $\Lambda$}\label{recon}

The arbitrary function $f(X)$ in expression (\ref{DL2}) is known as the 
{\it nonlocal distortion function}. Many models of dark energy and modified 
gravity have such free functions. The {\it Reconstruction Problem} consists 
of adjusting these functions to reproduce a given expansion history, which is
usually $\Lambda$CDM. The geometry is homogeneous, isotropic and spatially flat,
\begin{equation}
ds^2 = -dt^2 + a^2(t) d\vec{x} \!\cdot\! d\vec{x} \qquad \Longrightarrow
\qquad H(t) \equiv \frac{\dot{a}}{a} \; . \label{FLRW}
\end{equation}
For the reconstruction problem we assume $a(t)$ and all its derivatives 
are known functions of time. 

It is straightforward to solve the reconstruction problem for scalar 
quint\-es\-sence models \cite{TW1,scalrecon,RPW} and a brief presentation of 
the solution will help fix ideas. For the geometry (\ref{FLRW}) the scalar
depends just on time and only two of Einstein's equations are nontrivial,
\begin{eqnarray}
3 H^2 & = & 8\pi G \Bigl[ \frac12 \dot{\varphi}^2 + V(\varphi)\Bigr] \; ,
\label{E1} \\
-2 \dot{H} - 3 H^2 & = & 8\pi G \Bigl[ \frac12 \dot{\varphi}^2 - 
V(\varphi)\Bigr] \; . \label{E2}
\end{eqnarray}
As usual, $G$ is the Newton constant, and a dot denotes the time derivative.
By adding (\ref{E1}) and (\ref{E2}) one obtains the relation,
\begin{equation} \nonumber
-2\dot{H} = 8 \pi G \, \dot{\varphi}^2 \; .
\end{equation}
Hence one can reconstruct the scalar's evolution provided the Hubble parameter is 
monotonically decreasing, 
\begin{equation}
\varphi(t) = \varphi_0 \pm \int_0^{t} \!\! dt' \sqrt{\frac{-2 \dot{H}(t')}{
8\pi G}} \qquad \Longleftrightarrow \qquad t = t(\varphi) \; .
\end{equation}
This relation can be inverted (numerically if need be) to give the time as a
function of the scalar, $t(\varphi)$. One then determines the potential by 
subtracting (\ref{E2}) from (\ref{E1}) and evaluating the resulting function of 
time at $t = t(\varphi)$,
\begin{equation}
V(\varphi) = \frac1{8 \pi G} \Biggl[ \dot{H}\Bigl(t(\varphi)\Bigr) + 
3 H^2\Bigl(t(\varphi)\Bigr) \Biggr] \; .
\end{equation}

I will review a similar construction for the nonlocal distortion function $f(X)$
that was devised with Cedric Deffayet \cite{Cedric}. It and like techniques
\cite{Tomi,Vernov} allow one to reproduce the $\Lambda$CDM expansion history --- 
or any other $a(t)$ without a cosmological constant or dark energy. This is an 
important distinction between nonlocal cosmology and $f(R)$ models \cite{nof(R)}.

For the geometry (\ref{FLRW}) the nontrivial field equations are,
\begin{eqnarray}
3H^2  + \Delta G_{00} &=& 8 \pi G \rho \; , \label{EQrho} \\
-2 \dot{H} - 3H^2  + \frac{1}{3 a^2} \delta^{ij} \Delta G_{ij} 
&=& 8 \pi G p \; . \label{EQP}
\end{eqnarray}
Here $\rho$ and $p$ are the energy density and pressure, which are assumed
to be known functions of time. For example, to reproduce the $\Lambda$CDM expansion
history without $\Lambda$ we would use,
\begin{eqnarray}
\rho(t) & = & \frac{3 H_0^2}{8 \pi G} \Bigl[ \Omega_{\rm mat} \Bigl( \frac{a_0}{a(t)}
\Bigr)^3 + \Omega_{\rm rad} \Bigl( \frac{a_0}{a(t)} \Bigr)^4 \Bigl] \; , \\ 
p(t) & = & \frac{3 H_0^2}{8 \pi G} \Bigl[ 0 + \frac13 \Omega_{\rm rad} \Bigl( 
\frac{a_0}{a(t)} \Bigr)^4 \Bigl] \; ,
\end{eqnarray}
where $a_0$ and $H_0$ are the current values of the scale factor and the Hubble 
parameter, and $\Omega_{\rm mat}$ and $\Omega_{\rm rad}$ are the fractions of the 
current $\Lambda$CDM energy density in matter and radiation. The specialization of 
the nonlocal modification terms to the geometry (\ref{FLRW}) gives,
\begin{eqnarray}
\lefteqn{\Delta G_{00} = \Bigl[3 H^2 + 3 H \partial_{t}\Bigr] 
\Biggl\{ f \Bigl(\frac1{\square} R\Bigr) + \frac1{\square} 
\Biggl[ R f'\Bigl(\frac1{\square} R\Bigr)\Biggr] \Bigg\} } \nonumber \\
& & \hspace{4cm} + \frac12 \partial_{t} \Bigl(\frac1{\square} R\Bigr) \times 
\partial_{t} \Biggl(\frac1{\square} \Biggl[ R f'\Bigl(\frac1{\square} R 
\Bigr) \Biggr] \Biggr) , \qquad \label{D200} \\
\lefteqn{\Delta G_{ij} = -\Bigl[2 \dot{H} + 3 H^2 + 2 H \partial_{t} + 
\partial_{t}^2\Bigr] \Biggl\{ f\Bigl(\frac1{\square} R\Bigr) + 
\frac1{\square} \Biggl[ R f'\Bigl(\frac1{\square} R\Bigr)\Biggr] \Bigg\} 
g_{ij} } \nonumber \\
& & \hspace{4cm} + \frac12 \partial_{t} \Bigl(\frac1{\square} R\Bigr) \times 
\partial_{t} \Biggl(\frac1{\square} \Biggl[ R f'\Bigl(\frac1{\square} R 
\Bigr) \Biggr] \Biggr) g_{ij} \; . \qquad \label{D2ij}
\end{eqnarray}
For this geometry the Ricci scalar is $R(t) = 6\dot{H} + 12 H^2$ and the action of 
$\frac1{\square}$ on some function of time $W(t)$ can be expressed as a double integral
from the initial time $t_i$,
\begin{equation} \label{1BH}
\frac1{\square} \Bigl[W\Bigr](t) = -\int_{t_i}^t \!\! dt' \, \frac1{a^{3}(t')}
\int_{t_i}^{t'} \!\! dt'' \, a^3(t'') \, W(t'') \; . \label{invBox}
\end{equation}

\begin{figure*}
\includegraphics[width=\textwidth]{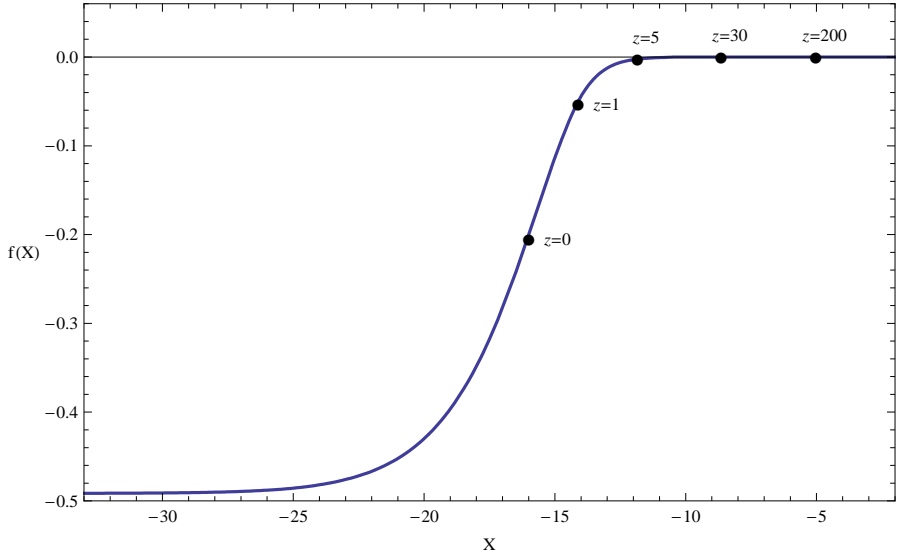}
\caption{Plot (solid blue curve) of the reconstructed function $f(X)$ for
the nonlocal cosmology reproducing $\Lambda$CDM background cosmological
evolution, with the same matter content but no cosmological constant. The
parameters corresponding to the background cosmology are those of the
five-year WMAP release \cite{WMAP5}. Circles indicate values of the function
$f(X)$ with the corresponding value of the redshift $z$ indicated above.}
\label{graph}
\end{figure*}

The first step of our construction is taking the difference between relations (\ref{EQrho}) 
and (\ref{EQP}). This leads to a simple equation for the function $F(t)$, 
\begin{equation}
F(t) \equiv f\Bigl( X(t) \Bigr) + \frac{1}{\square}\Bigl[ f'\Bigl( X(t) \Bigr) R(t)\Bigr]
\qquad , \qquad X(t) \equiv \frac1{\square}\Bigl[ R\Bigr](t) \; . \label{Ff2}
\end{equation}
This differential equation is, 
\begin{equation}
\ddot{F} + 5 H\dot{F} + (2 \dot{H} + 6 H^2) F = 
8 \pi G (\rho -  p) - (2\dot{H} + 6 H^2) \; . \label{ODE}
\end{equation}
Recognizing that $F = \frac1{a^2}$ is a homogeneous solution permits us to reduce the order
so that the solution can be expressed in terms of an integral. Changing the dependent 
variable to $\Phi(t) \equiv a^2(t) \times F(t)$ allows us to express (\ref{ODE}) as,
\begin{equation}
\frac{d}{dt} \Bigl[ a^3 \dot{\Phi} \Bigr] = a^5 \Bigl[ 8\pi G (\rho - p) - (2\dot{H} 
+ 6 H^2) \Bigr] \equiv a^3(t) \times S(t) \; . \label{ODE2}
\end{equation}
The solution to (\ref{ODE2}) for general $a(t)$ can be written as a double integral,
\begin{equation}
\Phi(t) = \Phi_i + \int_{t_i}^{t} \!\! dt' \frac{a_1^3 \dot{\Phi}_1}{a^3(t)} +
\int_{t_i}^{t} \!\! dt' \! \int_{t_i}^{t'} \!\! dt'' \, a^3(t'') S(t'') \; . \label{Phi}
\end{equation}
Requiring no deviation from general relativity at the initial time corresponds to the
initial values $\Phi_i = 0 = \dot{\Phi}_i$, which allows us to express the function
$F(t)$ using the notation (\ref{invBox}),
\begin{equation}
F(t) = -\frac1{a^2(t)} \frac1{\square} \Bigl[ S\Bigr](t) \; . \label{Fsol}
\end{equation}

The next step is to act $\square \rightarrow -(\partial_t^2 + 3 H \partial_t)$ on 
relation (\ref{Ff2}) and regard the result as a differential equation for the
function $\mathcal{F}(t) \equiv f(X(t))$,
\begin{equation}
\ddot{\mathcal{F}} + (3 H - R \dot{X}) \dot{\mathcal{F}} = \ddot{F} + 3 H \dot{F} 
\; . \label{ODE3}
\end{equation}
We can find an integrating factor for (\ref{ODE3}) in terms of the function,
\begin{equation}
Z(t) \equiv \int_{t_i}^{t} \!\! dt' \frac{R(t')}{a^3(t')} \! \int_{t_i}^{t'} \!\!
dt'' \, a^3(t'') R(t'') \; . \label{Zdef}
\end{equation}
Multiplying (\ref{ODE3}) by $a^3 e^{Z}$ gives a form which can be immediately integrated,
\begin{equation}
\frac{d}{dt} \Biggl[ a^3(t) \, e^{Z(t)} \, \dot{\mathcal{F}}(t) \Biggr] = e^{Z(t)} 
\frac{d}{dt} \Bigl[ a^3(t) \, F(t) \Bigr] \; .
\end{equation}

The penultimate step is to solve for time as a function of $X$,
\begin{equation}
X(t) = -\int_{t_i}^{t} \!\! dt' \frac1{a^3(t')} \! \int_{t_i}^{t'} \!\! dt'' \,
a^3(t'') \, R(t'') \qquad \Longleftrightarrow \qquad t(X) \; . \label{Xdef}
\end{equation}
The desired nonlocal distortion function is $\mathcal{F}(t)$ evaluated at $t = 
t(X)$,
\begin{equation}
f(X) = \mathcal{F}\Bigl( t(X)\Bigr) \; . 
\end{equation}
All of these steps are straightforward to implement numerically for any scale
factor $a(t)$. Fig \ref{graph} shows the nonlocal distortion function that 
Deffayet and I \cite{Cedric} constructed for the $\Lambda$CDM values pertinent to 
the Five-Year WMAP data \cite{WMAP5}. (Dodelson and Park have carried out the 
procedure with current data \cite{Scott}.) There is a simple analytic form for 
$f(X)$ which is indistinguishable from the numerical solution \cite{Cedric},
\begin{equation}
f(X) \approx 0.245 \Bigl[\tanh\Bigl(0.350 Y \!+\! 0.032 Y^2 \!+\!
0.003 Y^3\Bigr) \!-\! 1\Bigr] \quad , \quad Y \equiv X + 16.5 \; .
\label{analytic}
\end{equation}
In fact the current data on the expansion history do not require keeping the 
higher powers of $Y$ in the argument of the $\tanh$ function.

\section{Perfect Screening for Free}\label{screen}

Modified gravity models based on changing $R$ to $f(R)$ suffer from a 
major problem in that $R$ typically has the same sign for cosmology, 
where we want big effects to explain the acceleration data, and for the 
solar system, where significant deviations from general relativity are 
not permissible. This has prompted the development of elaborate 
``chameleon mechanisms'' in which the extra scalar degree of freedom 
present in $f(R)$ models is light in cosmological settings and heavy 
inside the solar system \cite{Justin}. Nonlocal cosmology differs from 
$f(R)$ models in two crucial ways:
\begin{itemize}
\item{There are no extra degrees of freedom to mediate new forces; and}
\item{The factor of $\square^{-1}$ acting on $R$ allows us to define
the nonlocal distortion function so that there are no changes from
general relativity {\it at all} in a gravitationally bound system,
without affecting cosmology.}
\end{itemize}
The first point will be demonstrated in section~\ref{DoF}; it
is the second point which concerns us here.

The key fact is that the scalar d'Alembertian $\square \equiv
(-g)^{-\frac12} \partial_{\mu} (\sqrt{-g} \, g^{\mu\nu}
\partial_{\nu})$ has different signs when acting on functions of
time than on functions of space. This is obvious from the flat space
limit $\square \rightarrow -\partial_t^2 + \nabla^2$. In the background 
cosmology, and small perturbations about it, the time dependence of the 
Ricci scalar is stronger than its dependence upon space. This means that
$\frac1{\square} R$ is typically negative for cosmology. That is clear 
from the minus sign in expression (\ref{Xdef}). Indeed, the reconstruction
procedure of section \ref{recon} determines the nonlocal distortion 
function $f(X)$ only for $X < 0$. This can be seen from the graph in
Fig. \ref{graph}: reconstruction does not fix the form of $f(X)$ for
$X > 0$.

Although gravitationally bound systems are not always static, it is
generally true that their dependence upon space is stronger than
their dependence upon time. That means $\frac1{\square} R$ is positive
inside a gravitationally bound system. Further, reproducing the
$\Lambda$CDM expansion history requires $f(0) = 0$ \cite{Cedric}. If
we wish to completely eliminate any corrections inside gravitationally
bound systems it will suffice to define $f(X) = 0$ for all $X > 0$.
Hence there is a very simple way to make screening 100\% effective
inside the solar system, the galaxy, or any other gravitationally
bound system, all without affecting the model's behavior for
cosmology.

Some model builders consider that requiring $f(X)$ to vanish for $X > 0$
represents a horrendous amount of fine-tuning. This objection would be tenable
if one regarded nonlocal cosmology as some sort of fundamental theory, but 
that is precisely what it cannot be. The initial time $t_i$ we saw in 
expressions (\ref{invBox}), (\ref{Phi}), (\ref{Zdef}) and (\ref{Xdef}) has 
no place in a fundamental theory. If the model is to avoid Newton's famous 
philosophical injunction \cite{IN} --- and more modern pitfalls \cite{EW} 
--- its nonlocal modifications must represent the most cosmologically 
significant quantum gravitational corrections to the effective field equations 
from the vast ensemble of infrared gravitons produced during primordial inflation. 
Those gravitons were of {\it horizon scale} when they were ripped from the 
vacuum. (That is why one can study the process using quantum general relativity.) 
It makes perfect sense for them to have the biggest effect on very large scales, 
and little or no effect on small scales. This is not exquisite fine-tuning but 
rather an inevitable and completely natural consequence of the putative physical 
origin of nonlocal cosmology.

\section{Degrees of Freedom and Kinetic Stability}\label{DoF}

The past two sections have shown how the nonlocal distortion function can be 
defined to reproduce the $\Lambda$CDM expansion history, without changing the 
predictions of general relativity for gravitationally bound systems. We seek 
now to understand what sort of initial value formalism the model possesses. 
In particular, what are its degrees of freedom and constraints? And do any of
its degrees of freedom ever become ghosts? These issues were settled by a recent
paper written with Stanley Deser \cite{opus2}.  

\subsection{The Local Version}\label{local}

Shortly after the nonlocal cosmology model (\ref{DL2}) was proposed Nojiri and 
Odinstov devised a local version based on two scalar fields \cite{Sergei}. Their 
idea is for  variation with respect to a scalar $\xi$ to enforce the equation
$\square \phi = R$, so that $\phi$ becomes $\frac1{\square} R$ plus an
arbitrary homogeneous solution. The nonlocal distortion terms in
(\ref{DL2}) are subject to the modification,
\begin{equation}
R f\Bigl( \frac1{\square} R \Bigr) \sqrt{-g} \longrightarrow R f(\phi)
\sqrt{-g} + \xi \Bigl(\square \phi \!-\! R\Bigr) \sqrt{-g} \; .
\label{locver}
\end{equation}
This localized version of the model contains two scalar degrees of
freedom which are not present in the original, nonlocal model
(\ref{DL2}) \cite{Kosh}. The purpose of this subsection is
to demonstrate that one of these extra degrees of freedom is a
ghost, as pointed out to me by Gilles Esposito-Farese.

To exhibit the ghost, first partially integrate the scalar kinetic term of 
(\ref{locver}). Next express this as a difference of two squares,
\begin{eqnarray}
\lefteqn{\xi \square \phi \sqrt{-g} \longrightarrow -\partial_{\mu}
\phi \partial_{\nu} \xi g^{\mu\nu} \sqrt{-g} \; , } \\
& & \hspace{-.3cm} = -\frac14 \partial_{\mu} \Bigl( \phi \!+\!
\xi\Bigr) \partial_{\nu} \Bigl( \phi \!+\! \xi\Bigr) g^{\mu\nu}
\sqrt{-g} + \frac14 \partial_{\mu} \Bigl( \phi \!-\! \xi\Bigr)
\partial_{\nu} \Bigl( \phi \!-\! \xi\Bigr) g^{\mu\nu} \sqrt{-g}
\; . \qquad \label{ghost}
\end{eqnarray}
With our spacelike signature one can see that the combination $\phi
+ \xi$ has positive kinetic energy whereas the combination $\phi -
\xi$ is a ghost.

The analysis I have just given is valid so long as the metric is any
nondynamical background. But our metric is of course dynamical ---
that is the point of modified gravity! --- and one might worry about 
differentiated metric fields from the $R f(\phi)$ term mixing with the 
other fields to stabilize the ghost. First, note that such a stabilization
would not be phenomenologically acceptable, if it did occur, because 
$\dot{\phi}(t,\vec{x}) - \dot{\xi}(t,\vec{x})$ would still develop large 
inhomogeneities because $\phi - \xi$ is a ghost in {\it any} background
metric. However, the question was analysed in detail by Nojiri, Odintsov,
Sasaki and Zhang \cite{NOSZ}, who concluded that a ghost would form 
unless,
\begin{equation}
6 f'(\phi) > 1 + f(\phi) - \xi > 0 \; . \label{cond}
\end{equation}
The authors asserted that classical evolution from special initial conditions
could result in these inequalities being preserved for long periods. That is
true enough, but it ignores the virulence of kinetic instabilities. Kinetic 
instabilities are driven by the infinite volume of phase space in the ultraviolet, 
and they will lead to instantaneous decay if any initial value data permit them.
The fact that condition (\ref{cond}) is {\it linear} in the independent field 
$\xi(t,\vec{x})$ means that the local model cannot be stable.  

It might be worried that the existence of this unstable scalar
extension of the original model ensures that nonlocal cosmology is
itself unstable. Of course the original, nonlocal model is just the
local one subject to the initial value constraints $\phi(t_i,\vec{x})
= \xi(t_i,\vec{x})= 0 = \dot{\phi}(t_i,\vec{x}) =
\dot{\xi}(t_i,\vec{x})$. Why, it might be wondered, can we not excite
the ghost degree of freedom? However, general relativity offers an
example in which constraining a particular variable converts an
unstable model into a stable one. The variable we have in mind is of
course the Newtonian potential, which would be a ghost, and fatal to
the theory, were it not a constrained field. So it is at least
conceivable that the original, nonlocal model
(\ref{DL2}) is stable. In the remainder of this section I will
demonstrate that it is at indeed free of ghosts.

\subsection{Synchronous gauge}\label{synchgauge}

Synchronous gauge is the frame of freely falling observers who are released 
from a spacelike surface with zero relative velocities \cite{Lifsh},
\begin{equation}
ds^2 = -dt^2 + h_{ij}(t,\vec{x}) dx^i dx^j \; . \label{synch}
\end{equation}
Note that $h_{ij}(t,\vec{x})$ denotes the full 3-metric, including
the cosmological background of $\delta_{ij} a^2(t)$. Synchronous gauge 
has well-known problems with caustics which should not be important 
for our purposes. We believe that the basic analysis and conclusions of 
this section would apply for any lapse and shift.

In synchronous gauge the covariant scalar d'Alembertian takes the
form,
\begin{equation}
\square = -\partial_{t}^2 - \frac12 h^{ij} \dot{h}_{ij} \partial_{t}
+ \frac1{\sqrt{h}} \, \partial_{i} \Bigl( \sqrt{h} \, h^{ij}
\partial_j \Bigr) \; . \label{dAlem}
\end{equation}
Here and henceforth, $h^{ij}$ denotes the inverse of the spatial
metric $h_{ij}$, $h$ stands for the determinant of $h_{ij}$, and an
overdot represents differentiation with respect to time. The various
curvatures we require are,
\begin{eqnarray}
R_{00} & = & -\frac12 h^{k\ell} \ddot{h}_{k\ell} + \frac14 h^{ik}
h^{j\ell} \dot{h}_{ij} \dot{h}_{k\ell} \; , \label{R00} \\
R_{ij} & = & \frac12 \ddot{h}_{ij} + \frac14 h^{k\ell} \dot{h}_{ij}
\dot{h}_{k\ell} - \frac12 h^{k\ell} \dot{h}_{ik} \dot{h}_{j\ell} +
\mbox{}^{(3)}R_{ij} \; , \label{Rij} \qquad \\
R & = & h^{k\ell} \ddot{h}_{k\ell} + \frac14 h^{ij} h^{k\ell}
\dot{h}_{ij} \dot{h}_{k\ell} - \frac34 h^{ik} h^{j\ell} \dot{h}_{ij}
\dot{h}_{k\ell} + \mbox{}^{(3)}R \; . \qquad \label{Ricci}
\end{eqnarray}
Note that we follow the usual convention in which a superscript
$(3)$ before a quantity denotes its specialization to the purely
spatial geometry.

\subsection{Initial value data and constraints}\label{initial}

Let us first see that the nonlocal field equations
require the same initial value data as general relativity, namely,
the values of the 3-metric and its first time derivative at $t = t_i$:
$h_{ij}(t_i,\vec{x})$ and $\dot{h}_{ij}(t_i,\vec{x})$. The retarded
Green's function associated with $\frac1{\square}$ is defined by the
differential equation,
\begin{equation}
\sqrt{h} \, \square \, G[h](t,\vec{x};t',\vec{x}') = \delta(t \!-\!
t') \delta^3(\vec{x} \!-\! \vec{x}') \; , \label{Geqn}
\end{equation}
subject to retarded boundary conditions,
\begin{equation}
G[h](0,\vec{x};t',\vec{x}') = 0 \qquad \forall \, t' > t \; . \label{GBC}
\end{equation}
Even though we cannot solve equations (\ref{Geqn}-\ref{GBC}) for an
arbitrary 3-metric, their form clearly defines the Green's function
$G[h]$ at time $t$ using only the values of $h_{ij}$ and its first
time derivative for times less than or equal to $t$.

Because $\frac1{\square} R$ is the integral of
$G[h](t,\vec{x};t',\vec{x}')$ multiplied by the Ricci scalar, we
need only consider the second time derivatives of the $R$; the lower
time derivatives and all spatial derivatives are shielded by the
inverse differential operator. From expression (\ref{Ricci}) we see
that these second time derivatives can be written in form,
\begin{equation}
R = \partial_{t}^2 \ln(h) + \frac14 \Bigl( h^{ij} h^{k\ell} + h^{ik}
h^{j\ell} \Bigr) \dot{h}_{ij} \dot{h}_{k\ell} + \mbox{}^{(3)}R \; .
\label{Rform}
\end{equation}
Now use relation (\ref{dAlem}) to express second time derivatives in
terms of $\square$,
\begin{equation}
\partial_{t}^2 = -\square - \frac12 h^{ij} \dot{h}_{ij} \partial_t +
\frac1{\sqrt{h}} \, \partial_i \Bigl( \sqrt{h} h^{ij} \partial_j
\Bigr) \; . \label{dtform}
\end{equation}
We can obviously combine relation (\ref{dtform}) with (\ref{Rform})
to conclude,
\begin{eqnarray}
\lefteqn{R = -\square \ln(h) + \frac14 \Bigl( h^{ik} h^{j\ell} \!-\!
h^{ij} h^{k\ell} \Bigr) \dot{h}_{ij} \dot{h}_{k\ell} } \nonumber \\
& & \hspace{2.5cm} + h^{ij} \Bigl( \Gamma^k_{~ ij , k} \!+\!
\Gamma^k_{~ k i , j} \!-\! \Gamma^k_{~ k\ell} \Gamma^{\ell}_{~ ij}
\!-\! \Gamma^{k}_{~ \ell i} \Gamma^{\ell}_{~ kj} \!-\! \Gamma^k_{~
ki} \Gamma^{\ell}_{~ \ell j} \Bigr) \; . \qquad \label{Rbox}
\end{eqnarray}
Here $\Gamma^{k}_{~ij} \equiv \frac12 h^{k\ell} (h_{\ell i , j} +
h_{j\ell , i} - h_{ij , \ell})$ is the spatial affine connection and
a comma denotes partial differentiation.

With relations (\ref{Geqn}-\ref{GBC}), equation (\ref{Rbox}) shows
that $\frac1{\square} R$ involves only the usual initial value data of
general relativity: $h_{ij}(0,t_i,\vec{x})$ and
$\dot{h}_{ij}(t_i,\vec{x})$. We can show that these initial value data
are apportioned the same way (as general relativity) between
constrained fields and gravitational radiation by examining the
nonlocal corrections $\Delta G_{00}$ and $\Delta G_{0i}$ to the
constraint equations. Note first from (\ref{GBC}) that
$\frac1{\square}$ and its first time derivative both vanish at $t = t_i$.
Further, the nonlocal distortion function vanishes at $X = 0$. So we
need only examine the two terms of $\Delta G_{\mu0}$ in which two
covariant derivatives act upon $f(\frac1{\square} R) + \frac1{\square} 
[R f'(\frac1{\square} R)]$. It is simple to show that neither of the two
combinations in the constraint equations contains a second time
derivative,
\begin{eqnarray}
g_{00} \square - D_0 D_0 & = & \frac12 h^{k\ell} \dot{h}_{k\ell}
\partial_{t} - \frac1{\sqrt{h}} \, \partial_k \Bigl( \sqrt{h} \,
h^{k \ell} \partial_{\ell} \Bigr) \; , \qquad \\
g_{0i} \square - D_0 D_i & = & -\partial_{t} \partial_i + \frac12
h^{k\ell} \dot{h}_{i k} \partial_{\ell} \; .
\end{eqnarray}
Hence we conclude that the nonlocal corrections to the constraint
equations vanish at $t = t_i$,
\begin{equation}
t = t_i \qquad \Longrightarrow \qquad \Delta G_{00} = 0 = \Delta G_{0i} \; .
\end{equation}
This means that the nonlocal model possesses the same initial value
data as general relativity, and that this initial value data is
subject to precisely the same constraints as in general relativity.

\subsection{No ghosts}\label{ghosts}

To see that there are no ghosts it suffices to examine the second time
derivative terms (in synchronous gauge) of the dynamical equations,
$G_{ij} + \Delta G_{ij} = 8\pi G T_{ij}$. The second derivatives of
$h_{ij}(t,\vec{x})$ in the Einstein tensor follow from expressions
(\ref{Rij}-\ref{Ricci}),
\begin{equation}
G_{ij} = \frac12 \ddot{h}_{ij} - \frac12 h_{ij} h^{k\ell}
\ddot{h}_{k\ell} + O(\partial_t) \; . \label{GRkey}
\end{equation}
Of course it is only the first term, $\frac12 \ddot{h}_{ij}$, which
deals with unconstrained fields; the second term represents
completely constrained fields. Because general relativity has no
ghosts, we need only check that the nonlocal corrections in
(\ref{mod2}) don't change the sign of the $\frac12 \ddot{h}_{ij}$
term in (\ref{GRkey}).

The work of the previous subsection shows that local second time
derivatives can only come from the parts of $\Delta G_{ij}$ which
either multiply $G_{ij}$ or have two covariant derivatives acting
upon $f(\frac1{\square} R) + \frac1{\square} [R f'(\frac1{\square} 
R)]$. The latter terms are simple to analyze,
\begin{equation}
g_{ij} \square - D_i D_j = h_{ij} \square + O(\partial_t) \; .
\end{equation}
The local second derivative terms are therefore,
\begin{eqnarray}
\lefteqn{G_{ij} + \Delta G_{ij} = \frac12 \ddot{h}_{ij} \times
\Biggl[ 1 + f\Bigl( \frac1{\square} R\Bigr) + \frac1{\square} \Bigl[ R
f'\Bigl( \frac1{\square} R\Bigr) \Bigr] \Biggr] } \nonumber \\
& & \hspace{-.5cm} - \frac12 h_{ij} h^{k\ell} \ddot{h}_{k\ell}
\times \Biggl[ 1 + f\Bigl( \frac1{\square} R\Bigr) + \frac1{\square} 
\Bigl[R f'\Bigl( \frac1{\square} R\Bigr) \Bigr] - 4 f'\Bigl( 
\frac1{\square} R\Bigr) \Biggr] + O(\partial_t) \; . \qquad \label{nonkey}
\end{eqnarray}
Only the first line of expression (\ref{nonkey}) represents the
unconstrained, dynamical part of $h_{ij}$. By comparing with the
approximate analytic form (\ref{analytic}) of the nonlocal
distortion function $f(X)$ we see that the coefficient of the
dynamical term is reduced at late times, but never by enough to make
it change sign. We therefore conclude that no dynamical graviton
mode ever becomes a ghost.

\section{The Negative Verdict of Structure Formation}\label{struc}

Midway through the reconstruction work with Cedric Deffayet \cite{Cedric} 
it became obvious to me that models of the type (\ref{DL2}) were likely to 
have problems with structure formation because of how they contrive to 
support late time acceleration without dark energy. To see the problem, 
specialize the Friedmann equation of general relativity to an energy density 
comprised of only nonrelativistic matter,
\begin{equation}
3 H^2 = 8 \pi G \rho \longrightarrow 8 \pi G \rho_0 \Bigl( \frac{a_0}{a(t)} 
\Bigr)^3 \; . \label{Friedmann}
\end{equation}
Dark energy was posited because the data says the left hand side of 
(\ref{Friedmann}) is approaching a constant whereas the right hand side is
falling off. To understand how nonlocal cosmology resolves this conundrum it
suffices to specialize the nonlocal corrections (\ref{mod2}) to slowly varying 
and nearly homogeneous geometries. Under these two assumptions one can drop 
all the derivatives,
\begin{equation}
\Delta G_{\mu\nu} \approx \Bigl[ R_{\mu\nu} - \frac12 g_{\mu\nu} R\Bigr]
\Biggl\{ f\Bigl( \frac1{\square} R \Bigr) + \frac1{\square} \Biggl[ R
f'\Bigl( \frac1{\square} R \Bigr) \Biggr] \Biggr\} \; . \label{simpDG}
\end{equation}
Of course expression (\ref{simpDG}) is proportional to the Einstein tensor,
which means that, for slowly varying and nearly homogeneous geometries, we 
can regard the modified gravity equations as those of general relativity with
a time-varying effective Newton constant,
\begin{equation}
G_{\rm eff} \equiv \frac{G}{1 + f(\frac1{\square} R) + \frac1{\square} [R
f'(\frac1{\square} R)]} \; . \label{Geff}
\end{equation}
The right hand side of the modified Friedmann equation ($3 H^2 \approx 8\pi G_{\rm eff}
\rho$) is brought into balance with the nearly constant left hand side because
the growth in $G_{\rm eff}$ compensates for the redshift of the energy density.

{\it Strengthening the force of gravity is dangerous!} The fact that it happens 
for models of the form (\ref{DL2}) convinced me, back in the fall of 2008, that 
this class of models cannot be correct. However, the problem is not as bad as one 
might imagine. For example, there is nothing wrong inside the solar system where 
we have very strong constraints on time variation in Newton's constant. For any 
gravitationally bound system the geometry is not homogeneous so the approximation 
(\ref{simpDG}) is invalid. In any case the effective Newton constant (\ref{Geff}) 
degenerates to $G$ inside any gravitationally bound system because $X \equiv 
\frac1{\square} R > 0$ and $f(X)$ vanishes for positive $X$, as we saw in section 
\ref{screen}. To encounter a real problem one must study the regime in which 
perturbations around the cosmological background are small. That is what Dodelson 
and Park did \cite{Park,Scott}. I won't attempt to reproduce their work, but I 
will explain some of the preliminaries and summarize their conclusions.

Dodelson and Park consider scalar perturbations to (\ref{FLRW}),
\begin{equation}
ds^2 = -\Bigl[1 + 2 \Psi(t,\vec{x})\Bigr] dt^2 + a^2(t) \Bigl[ 1 + 2 \Phi(t,\vec{x})
\Bigr] \; . \label{potentials}
\end{equation}
They treat the potentials $\Psi(t,\vec{x})$ and $\Phi(t,\vec{x})$ to be small spatial
plane waves,
\begin{equation}
\Phi(t,\vec{x}) = \widetilde{\Phi}(t,\vec{k}) e^{i \vec{k} \cdot \vec{x}} \qquad ,
\qquad \Psi(t,\vec{x}) = \widetilde{\Psi}(t,\vec{k}) e^{i \vec{k} \cdot \vec{x}} \; .
\end{equation}
They then work out the first order perturbation $\delta (\widetilde{G}^{\mu}_{\nu} +
\Delta \widetilde{G}^{\mu}_{~\nu}) = 8\pi G \delta \widetilde{T}^{\mu}_{~\nu}$ to the 
nonlocally modified equation for nonrelativistic matter,
\begin{equation}
\delta \widetilde{T}^0_{~ 0} = -\delta \widetilde{\rho} \qquad , \qquad 
\delta \widetilde{T}^i_{~ j} = 0 \; . \label{Tmunu}
\end{equation}
It helps to see the purely temporal and spatial parts of $\delta \widetilde{G}^{\mu}_{~\nu}$,
\begin{eqnarray}
\delta \widetilde{G}^0_{~0} & = & -\frac{2 k^2}{a^2} \widetilde{\Phi} - 6 H 
\dot{\widetilde{\Phi}} + 6 H^2 \widetilde{\Phi} \; , \label{G00} \\
\delta \widetilde{G}^i_{~ j} & = & \frac{(k^i k^j \!-\! k^2 \delta^{ij})}{a^2} 
\Bigl[ \widetilde{\Phi} + \widetilde{\Psi} \Bigr] -2 \delta^{ij} \Bigl[
\ddot{\widetilde{\Phi}} + 3 H \dot{\widetilde{\Phi}} + H \dot{\widetilde{\Psi}}
- (2\dot{H} + 3 H^2) \widetilde{\Psi} \Bigr] . \qquad \label{Gij}
\end{eqnarray}
Because expression (\ref{Gij}) contains two independent tensor factors, we can infer 
three equations for the three variables $\widetilde{\Phi}(t,\vec{k})$, 
$\widetilde{\Psi}(t,\vec{k})$ and $\delta \widetilde{\rho}(t,\vec{k})$. 
The nonlocal corrections do not change this overall structure, although they do alter
important things such as the general relativistic relation $\widetilde{\Phi}(t,\vec{k})
= -\widetilde{\Psi}(t,\vec{k})$ between the two potentials.

I do wish to explain how to perturb functions of $X \equiv \frac1{\square} R$ \cite{TWd,Park},
\begin{equation}
\delta \widetilde{f(X)} = f'\Bigl(\overline{X}\Bigr) \!\times\! \delta \widetilde{X} 
\;\; , \;\; \overline{X}(t) = -\!\!\int_{t_i}^t \!\!\!\! \frac{dt'}{a^3(t')} \! \int_{t_i}^{t'} 
\!\!\!\!\! dt'' \, a^3(t'') \Bigl[ 6 \dot{H}(t'') \!+\! 12 H^2(t'') \Bigr] , \label{key1}
\end{equation}
where,
\begin{equation}
\delta \widetilde{X}(t,\vec{k}) = \!\! \int_{t_i}^t \!\!\! dt' \, G(t,t';k) 
\Biggl[ \frac{k^2}{a^2} \Bigl(4 \widetilde{\Phi} \!+\! 2 \widetilde{\Psi} \Bigr) \!+\! 
6 \ddot{\widetilde{\Phi}} \!+\! 6 H (4 \dot{\widetilde{\Phi}} \!-\! \dot{\widetilde{\Psi}}) 
\!+\! \dot{\overline{X}} (3 \dot{\widetilde{\Phi}} \!-\! \dot{\widetilde{\Psi}}) \Biggr] . 
\label{key2}
\end{equation}
The retarded Green's function,
\begin{equation}
G(t,t',k) \equiv -i \theta(t - t') a^3(t') \Bigl[ u(t,k) u^*(t',k) - u^*(t,k) u(t',k)\Bigr] \; ,
\end{equation}
is formed from the mode functions of the massless, minimally coupled scalar,
\begin{equation}
\ddot{u} + 3 H \dot{u} + \frac{k^2}{a^2} \, u = 0 \qquad {\rm and} \qquad 
u \dot{u}^* - \dot{u} u^* = \frac{i}{a^3} \; . \label{mode}
\end{equation}
For the subhorizon modes of interest, Dodelson and Park were able to ignore the
various time derivatives in expression (\ref{key2}) and they employed the WKB approximation 
for the retarded Green's function \cite{Park},
\begin{equation}
G(t,t';k) \approx -\frac{ \theta(t \!-\! t') a^2(t')}{k a(t)} \, \sin\Biggl[ k \! \int_{t'}^t 
\!\! \frac{dt''}{a(t'')} \Biggr] \; .
\end{equation}

Dodelson and Park found only small changes in the potential $\widetilde{\Phi}(t,\vec{k})$,
however, they discovered that the Newtonian potential $\widetilde{\Psi}(t,\vec{k})$ grows 
rapidly from about redshift $z = 1.5$ until it is nearly double that of general 
relativity at the current time \cite{Scott}. This is what one expects from a model
which strengthens the gravitational force, but it is a phenomenological disaster: 
weak lensing favors general relativity over nonlocal cosmology by almost $6 \sigma$,
and redshift space distortions favor general relativity by almost $8 \sigma$ \cite{Scott}.

\section{The Outlook for Nonlocal Modifications of Gravity}\label{outlook}

Particle theorists are notorious for concocting rosy interpretations of the
most devastating phenomenology. At the risk of further damaging our already
abysmal reputation, let me say that the recent paper by Dodelson and Park 
\cite{Scott} does not depress me, nor does it particularly surprise me. 
Their excellent work rules out the simplest class of models (\ref{DL2}), 
but I ceased taking those models seriously after realizing that they achieve 
cosmic acceleration by making the effective Newton constant grow. What one 
wants is a nonlocal model in which the cosmological constant becomes time 
dependent, and I eventually succeeded in devising a class of them in 
collaboration with Nick Tsamis \cite{TW3}. It seems clear that an elaboration 
of these models can reproduce the $\Lambda$CDM expansion history without 
disastrously enhancing early structure formation the way the simple models 
(\ref{DL2}) do. In fact the paper by Dodelson and Park encourages me because 
their larger message is that the latest data slightly favor a universe which 
is initially more uniform than the one predicted by general relativity 
\cite{Scott}. So this is a wonderful opportunity to build modified gravity 
models which match the data {\it better} than general relativity, and the 
Dodelson-Park paper provides crucial information on how to do it.

The class of models which I seek will involve a nonlocal invariant other than 
$X[g](x) \equiv \frac1{\square} R$. To understand why, I must recount that
Tsamis and I have proposed that the bare cosmological constant $\Lambda$ is 
not small (so there is no cosmological constant problem) but is today being 
screened by the self-gravitation between gravitons which were produced during 
a long epoch of primordial inflation \cite{TWa}. In the absence of a direct
computation from quantum gravity, we attempt to describe the screening process 
through a general class of nonlocal effective field equations --- characterized 
by another free function $f(-G \Lambda X)$ --- which interpolate between the 
known secular dependence of perturbative corrections \cite{TW2} to the 
conjectured regime in which quantum gravitational back-reaction becomes 
nonperturbatively strong. As long as our function $f(-G \Lambda X)$ grows 
monotonically and without bound, inflation ends in a characteristic way, and 
the nonlocal correction terms thereafter settle down to a constant, negative 
vacuum energy during radiation domination \cite{TW3}.

All of that sounds wonderful, and it is. However, our models respond exactly 
the wrong way to the onset of matter domination \cite{TW4}. Instead of 
engendering a phase of late time acceleration, our nonlocal corrections drive 
the total vacuum energy negative! This is because the Ricci scalar,
\begin{equation}
R \longrightarrow 6 \dot{H} + 12 H^2 = 6 (2 - \epsilon) H^2 \; ,
\end{equation}
has the same sign for vacuum energy domination ($\epsilon = 0$) and
matter domination ($\epsilon = \frac32$). So generating an ever-more negative 
vacuum energy during primordial inflation --- which we need to end inflation
--- means that the same thing also happens after the transition from radiation
domination ($\epsilon = 2$) to matter domination. 

The cure Tsamis and I proposed \cite{TW4} is to transform the factor of $-\Lambda$ 
in the function $f(-G \Lambda X)$, into a curvature invariant which changes sign 
when the slow roll parameter $\epsilon \equiv -\frac{\dot{H}}{H^2}$ varies from 
acceleration ($\epsilon < 1$) to deceleration ($\epsilon > 1$). When specialized
to the cosmological background (\ref{FLRW}) a reasonable choice for us is $R_{00}$,
\begin{equation}
R_{00} \longrightarrow -3(\dot{H} + H^2) = -3 (1 - \epsilon) H^2 \; .
\end{equation}
The problem is finding an invariant that becomes $R_{00}$ when specialized to the
cosmological background (\ref{FLRW}). The solution Tsamis and I proposed is to 
construct a timelike 4-velocity field $u^{\mu}[g](x)$ by normalizing the gradient
of $X[g](x) \equiv \frac1{\square} R$ \cite{TW4},
\begin{equation}
u^{\mu}[g](x) \equiv \frac{g^{\mu\nu}(x) \partial_{\nu} X[g](x)}{\sqrt{
-g^{\alpha\beta}(x) \partial_{\alpha} X[g](x) \partial_{\beta} X[g](x)}} \; .
\end{equation}
The desired invariant is obtained by contracting two factors of this into the 
Ricci tensor. So our new model is based on the replacement \cite{TW4},
\begin{equation}
f\Biggl(-G\Lambda \frac1{\square}\Bigl[R\Bigr] \Biggr) \longrightarrow
f\Biggl(G \frac1{\square}\Bigl[R \, u^{\mu} u^{\nu} R_{\mu\nu}\Bigr] 
\Biggr) \; .
\end{equation}
There is one more complication to get an acceptable model but that need not
concern us here. The main point I want to make is that many features of the
extended class of models will carry over from what has already been done for
(\ref{DL2}), so we do not face the task of beginning anew but rather that of
building on the existing foundations.

A final point is that I have also encountered the need for a second invariant in 
devising relativistic, metric-based extensions of Milgrom's modified gravity 
Modified Newtonian Dynamics (MOND)\cite{Moti}. MOND is impressively successful at 
explaining the structures of galaxies and galactic clusters without dark matter 
\cite{Bob}, so one suspects that it might represent the Newtonian limit of some 
fully relativistic, modified gravity model. One can build a nonlocal model based 
on $X \equiv \frac1{\square} R$ which reproduces the MOND force law\cite{SW} --- 
known as the Tully-Fisher relation --- but no such model can explain the observed 
amount of weak lensing \cite{DEW}. With Cedric Deffayet and Gilles Esposito-Farese 
I constructed a class of nonlocal models which accomplish both things by involving 
another invariant \cite{DEW},
\begin{equation}
Y[g](x) \equiv 4 g^{\mu\nu} \Biggl( \partial_{\mu} \Bigl[ u^{\alpha} u^{\beta} 
R_{\alpha\beta}\Bigr] \Biggr) \Biggl( \partial_{\nu} \Bigl[ u^{\rho} u^{\sigma} 
R_{\rho\sigma}\Bigr] \Biggr) \; .
\end{equation} 
One can hardly fail to recognize the analogy,
\begin{eqnarray}
{\rm Expansion\ History} & \Longleftrightarrow & {\rm Tully-Fisher} \; , 
\label{simple} \\
{\rm Structure\ Formation} & \Longleftrightarrow & {\rm Weak\ Lensing} \; .
\label{extension}
\end{eqnarray}
Modified gravity models involving only the nonlocal invariant $X[g] \equiv \frac1{\square} 
R$ can meet the requirements of line (\ref{simple}) without dark energy (left) or
dark matter (right). However, the requirements of the second line in each case 
necessitate a second invariant based upon $u^{\mu} u^{\nu} R_{\mu\nu}$. Perhaps there 
is a grand synthesis in which a single nonlocal gravity model describes both 
cosmology and gravitational forces on large scales, without the need for either dark 
energy or dark matter?  

\begin{acknowledgements}
It is a pleasure to thank my distinguished colleagues and collaborators: 
C. Deffayet, S. Deser, S. Dodelson, G. Esposito-Farese, S. Park and N. C. Tsamis.
\end{acknowledgements}


\begin{thebibliography}{}

\bibitem{SNIA} A. G. Riess, et al., Astron. J. {\bf 116} (1998) 1009,
astro-ph/9805201; S. Perlmutter, et al., Astrophys. J. {\bf 517}
(1999) 565, astro-ph/9812133.

\bibitem{newSN} M. Hicken et al., Astrophys. J. {\bf 700} (2009)
1097, arXiv:0901.4804.

\bibitem{data} WMAP Collabration, arXiv:1212.5226; Z. Hou et al.,
arXiv:1212.6267; J. L. Sievers et al., arXiv:1301.08024; Planck
Collaboration,\hfil\break arxiv:1303.5076.

\bibitem{reviews} S.'i. Nojiri and S. D. Odintsov, Phys. Rept.
{\bf 505} (2011) 59,\hfil\break arXiv:1011.0544; M. Li, X.-D. Li, S.
Wang and Y. Wang, Commun. Theor. Phys. {\bf 56} (2011) 525,
arXiv:1103.5870; K. Bamba, S. Capozziello, S.'i. Nojiri and S. D.
Odintsov, Astrophys. Space Sci. {\bf 342} (2012) 155,
arXiv:1205.3421.

\bibitem{Yun} Y. Wang and P. Mukherjee, Astrophys. J. {\bf 650}
(2006) 1, astro-ph/0604051; U. Alam, V. Sahni and A. A. Starobinsky,
JCAP {\bf 0702} (2007) 011, astro-ph/0612381.

\bibitem{Yunbook} Y. Wang, {\it Dark Energy} (Wiley-VCH, Weinheim,
Germany, 2010).

\bibitem{Lambdarevs} S. Weinberg, Rev. Mod. Phys. {\bf 61} (1989) 1;
V. Sahni and A. Starobinsky, Int. J. Mod. Phys. {\bf D9} (2000) 373,
astro-ph/9904398;
S. M. Carroll, Living Rev. Rel. {\bf 4} (2001) 1, astro-ph/0004075;
B. Ratra and P. J. E. Peebles, Rev. Mod. Phys. {\bf 75} (2003) 559,
astro-ph/0207347.

\bibitem{earlyquint} B. Ratra and P. J. E. Peebles, Phys. Rev. {\bf
D37} (1988) 3406; C. Wetterich, Nucl. Phys. {\bf B302} (1988) 668.

\bibitem{Paul} I. Zlatev, L.-M. Wang and P. J. Steinhardt, Phys. Rev.
Lett. {\bf 82} (1999) 896, astro-ph/9807002; P. J. Steinhardt, L.-M.
Wang and I. Zlatev, Phys. Rev. {\bf D59} (1999) 123504,
astro-ph/9812313; L.-M. Wang, R. R. Caldwell, J. P. Ostriker and P.
J. Steinhardt, Astrophys. J. {\bf 530} (2000) 17, astro-ph/9901388.

\bibitem{TW1} N. C. Tsamis and R. P. Woodard, Ann. Phys. {\bf 267} (1998) 145,
hep-ph/9712331.

\bibitem{scalrecon} T. D. Saini, S. Raychaudhury, V. Sahni and A. A.
Starobinsky, Phys. Rev. Lett. {\bf 85} (2000), astro-ph/9910231; S.
Capozziello, S. Nojiri and S. D. Odintsov, Phys. Lett. {\bf B634}
(2006) 93, hep-th/0512118; Z. K. Guo, N. Ohta and Y. Z. Zhang, Phys.
Rev. {\bf D72} (2005) 023504, astro-ph/0505253; Mod. Phys. Lett.
{\bf A22} (2007) 883, astro-ph/0603109.

\bibitem{Parker} L. Parker and A Raval, Phys. Rev. {\bf D60} (1999) 063512,
gr-qc/9905031; L. Parker and D. A. T. Vanzella, Phys. Rev. {\bf D69}
(2004) 104009, gr-qc/0312108.

\bibitem{Mark} S. M. Carroll, V. Duvvuri, M. Trodden and M. S. Turner,
Phys. Rev. {\bf D70} (2004) 043528, astro-ph/0306438.

\bibitem{NO} S. Nojiri and S. D. Odintsov, Int. J. Geom. Meth. Mod. Phys.
{\bf 4} (2007) 115, hep-th/0601213; 
W. Hu and I. Sawicki, Phys. Rev. {\bf D76} (2007) 064004, arXiv:0705.1158;
S. A. Appleby and R. A. Battye, Phys. Lett. {\bf B654} (2007) 7, arXiv:0705.3199;
A. A. Starobinsky, JETP Lett. {\bf 86} (2007) 157, arXiv:0706.2041. 

\bibitem{RPW} R. P. Woodard, Lect. Notes Phys. {\bf 720} (2007) 403,
astro-ph/0601672.

\bibitem{nof(R)} P. K. S. Dunsby, E. Elizalde, R. Goswami, S. Odintsov
and D. S. Gomez, Phys. Rev. {\bf D82} (2010) 023519,
arXiv:1005.2205.

\bibitem{nonloc} L. Parker and D. J. Toms, Phys. Rev. {\bf D32} (1985) 1409;
T. Banks, Nucl. Phys. {\bf B309} (1988) 493; C. Wetterich, Gen. Rel.
Grav. {\bf 30} (1998) 159, gr-qc/9704052;
A. O. Barvinsky, Phys. Lett. {\bf B572} (2003) 109, hep-th/0304229;
D. Espriu, T. Multamaki and E. C. Vagenas, Phys. Lett. {\bf B628}
(2005) 197, gr-qc/0503033; H. W. Hamber and R. M. Williams, Phys.
Rev. {\bf D72} (2005), 044026, hep-th/0507017; T. Biswas, A.
Mazumdar and W. Siegel, JCAP {\bf 0603} (2006) 009, hep-th/0508194;
D. Lopez Nacir and F. D. Mazzitelli, Phys. Rev. {\bf D75} (2007)
024003, hep-th/0610031; J. Khoury, Phys. Rev. {\bf D76} (2007)
123513, hep-th/0612052; S.~Capozziello, E.~Elizalde, S.~'i.~Nojiri
and S.~D.~Odintsov, Phys. Lett. {\bf B671} (2009) 193,
arXiv:0809.1535; T. Biswas, T. Koivisto and A. Mazumdar, JCAP {\bf
1011} (2010) 008, arXiv:1005.0590; Y.-l. Zhang and M. Sasaki, Int. 
J. Mod. Phys. {\bf D21} (2012) 1250006, arXiv:1108.2112; A. O. Barvinsky, 
Phys. Lett. {\bf b710} (2012) 12, arXiv:1107.1463; Phys. Rev. {\bf D85} 
(2012) 104018, arXiv:1112.4340; E. Elizalde, E. O. Pozdeeva and S. Y.
.Vernov, Phys. Rev. {\bf D85} (2012) 044002, arXiv:1110.5806; A. O.
Barvinsky and Y. V. Gusev, Phys. Part. Nucl. {\bf 44} (2013) 213,
arXiv:1209.3062; T. Biswas, A. Conroy, A. S. Koshelev and A. Mazumdar,
Class. Quant. Grav. {\bf 31} (2013) 015022 (2013), arXiv:1308.2319;
P. G. Ferreira and A. L. Maroto, Phys. Rev. {\bf D88} (2013) 123502,
arXiv:1310.1238; S. Foffa, M. Maggiore and E. Mitsou, arXiv:1311.3421; 
arXiv:1311.3435.

\bibitem{SW} M. E. Soussa and R. P. Woodard, Class. Quant. Grav. {\bf 20} 
(2003) 2737, astro-ph/0302030.

\bibitem{DEW} C. Deffayet, G. Esposito-Farese and R. P. Woodard, Phys. Rev. 
{\bf D84} (2011) 124054, arXiv:1106.4984.

\bibitem{IN} I. Newton, in {\it Four Letters from Sir Isaac Newton
to Doctor Bentley} (R. and J. Dodsley, Pall-Mall, 1756).

\bibitem{EW} D. A. Eliezer and R. P. Woodard, Nucl. Phys. {\bf B325} (1989)
389.

\bibitem{TWa} N. C. Tsamis and R. P. Woodard, Nucl. Phys. {\bf B474} (1996)
235, hep-ph/9602315; Int. J. Mod. Phys. {\bf D20} (2011) 2847, arXiv:1103.5134.

\bibitem{Alexei} A. A. Starobinsky, JETP Lett. {\bf 30} (1979) 682.

\bibitem{Sasha} A. M. Polyakov, Sov. Phys. Usp. {\bf 25} (1982) 187.

\bibitem{TWb} N. C. Tsamis and R. P. Woodard, Ann. Phys. {\bf 238} (1995) 1.

\bibitem{Steve} S. Weinberg, Phys. Rev. {\bf 140} (1965) B516.

\bibitem{Gabriele} G. Veneziano, Nucl. Phys. {\bf B44} (1972) 142.

\bibitem{TWc} N. C. Tsamis and R. P. Woodard, Ann. Phys. {\bf 253} (1997) 1,
hep-ph/9602316.

\bibitem{SY} A. A. Starobinsky and J. Yokoyama, Phys. Rev. {\bf D50}
(1994) 6357, astro-ph/9407016.

\bibitem{TW2} N. C. Tsamis and R. P. Woodard, Nucl. Phys. {\bf B724} (2005)
295, gr-qc/0505115; S. P. Miao and R. P. Woodard, Phys. Rev. {\bf
D74} (2006) 044019,gr-qc/0602110]; T. Prokopec, N. C. Tsamis and R.
P. Woodard, Annals Phys. {\bf 323} (2008) 1324, arXiv:0707.0847.

\bibitem{MW} S. P. Miao and R. P. Woodard, Class. Quant. Grav. {\bf 25} (2008)
145009, arXiv:0803.2377; H. Kitamoto and Y. Kitazawa, Phys. Rev.
{\bf D83} (2011) 104043, arXiv:1012.5930; Phys. Rev. {\bf D85}
(2012) 044062, arXiv:1109.4892.

\bibitem{ourmodel} S. Deser and R. P. Woodard, Phys. Rev. Lett. {\bf 99}
(2007) 111301, \hfill\break arXiv:0706.2151.

\bibitem{Yun2} E. Bertschinger, Astrophys. J. {\bf 648} (2006) 797, astro-ph/0604485;
W. Hu and I. Sawicki, Phys. Rev. {\bf 76} (2007) 104043, arXiv:0708.1190;
Y. Wang, JCAP {\bf 0805} (2008) 021, arXiv:0710.3885;
T. Baker, P. G. Perreira and C. skordis, Phys. Rev. {\bf D87} (2013) 024015, arXiv:1209.2117;
F. Simpson et al., arXiv:1212.3339;
D. Huterer et al., arXiv:1309.5385.

\bibitem{Tomi} T. Koivisto, Phys. Rev. {\bf D77} (2008) 123513, arXiv:0803.3399;
Phys. Rev. {\bf D78} (2008) 123505, arXiv:0807.3778.

\bibitem{Park} S. Park and S. Dodelson, Phys. Rev. {\bf D87} (2013)
024003,\hfil\break arXiv:1209.0836.

\bibitem{Scott} S. Dodelson and S. Park, arXiv:1310.4329.

\bibitem{SK} J. Schwinger, J. Math. Phys. {\bf 2} (1961) 407;
K. T. Mahanthappa, Phys. Rev. {\bf 126} (1962) 329;
P. M. Bakshi and K. T. Mahanthappa, J. Math. Phys. {\bf 4} (1963) 1;
J. Math. Phys. {\bf 4} (1963) 12;
L. V. Keldysh, Sov. Phys. JETP {\bf 20} (1965) 1018;
K. C. Chou, Z. B. Su, B. L. Hao and L. Yu, Phys. Rept. {\bf 118} (1985) 1;
R. D. Jordan, Phys. Rev. {\bf D33} (1986) 444;
E. Calzetta and B. L. Hu, Phys. Rev. {\bf D35} (1987) 495.

\bibitem{Cedric} C. Deffayet and R. P. Woodard, JCAP {\bf 0908} (2009)
023,\hfil\break arXiv:0904.0961.

\bibitem{Vernov} E. Elizalde, E. O. Pozdeeva and S. Yu Vernov,
Class. Quant. Grav. {\bf 30} (2013) 035002, arXiv:1209.5957; E.
Elizalde, E. O. Pozdeeva, S. Yu Vernov and Y.-l. Zhang,
arXiv:1302.4330.

\bibitem{WMAP5}  J. Dunkley et al., Astrophys. J. Suppl. {\bf 180} (2009) 
306, arXiv:0803.0586; E. Komatsu et al., Astrophys. J. Suppl. {\bf 180} 
(2009) 330, arXiv:0803.0547.
 
\bibitem{Justin} J. Khoury and A. Weltman, Phys. Rev. Lett. {\bf 93} (2004)
171104, astro-ph/0309300; Phys. Rev. {\bf D69} (2004) 044026,
astro-ph/0309411; P. Brax, C. van de Bruck, A.-C. Davis, J. Khoury
and A. Weltman, Phys. Rev. {\bf D70} (2004) 123518,
astro-ph/0408415.

\bibitem{opus2} S. Deser and R. P. Woodard, JCAP {\bf 1311} (2013) 036,
arXiv:1307.6639.

\bibitem{Sergei} S.'i. Nojiri  and S. D. Odintsov, Phys. Lett.
{\bf B659} (2008) 821,\hfil\break arXiv:0708.0924.

\bibitem{Kosh} N. A. Koshelev, Grav. Cosmol. {\bf 15} (2009) 220,
arXiv:0809.4927.

\bibitem{NOSZ} S.'i. Nojiri, S. D. Odintsov, M. Sasaki and Y.-l. Zhang,
Phys. Lett. {\bf B696} (2011) 278, arXiv:1010.5375.  

\bibitem{Lifsh} E. M. Lifshitz, J. Phys. USSR {\bf 10} (1946)
116.

\bibitem{SchYao} R. Schoen and S. T. Yao, Commun. Math. Phys. {\bf
65} (1979) 45; Commun. Math. Phys. {\bf 79} (1981) 231.

\bibitem{TWd} N. C. Tsamis and R. P. Woodard, Phys. Rev. {\bf D82} (2010)
063502, arXiv:1006.4834.

\bibitem{TW3} N. C. Tsamis and R. P. Woodard, Phys. Rev. {\bf D80} (2009) 
083512, arXiv:0904.2368; M. G. Romania, N. C. Tsamis and R. P. Woodard,
Lect. Notes Phys. {\bf 863} (2013) 375, arXiv:1204.6558.

\bibitem{TW4} N. C. Tsamis and R. P. Woodard, Phys. Rev. {\bf D81} (2010) 
103509, arXiv:1001.4929.

\bibitem{Moti} M. Milgrom, Astrophys. J. {\bf 270} (1983) 365, 371.

\bibitem{Bob} R. H. Sanders and S. S. McGaugh, Ann. Rev. Astron. Astrophys. 
{\bf 40} (2002) 263, astro-ph/0204521.

\end{thebibliography}
\end{document}